\documentclass[1p,12pt]{elsarticle}

\usepackage{axodraw}
\usepackage{graphicx}
\usepackage[latin1]{inputenc}
\usepackage[italian,english]{babel}
\usepackage{amsfonts,amsmath,amssymb,csquotes,hyperref}

\makeatletter
\def\ps@pprintTitle{%
	\let\@oddhead\@empty
	\let\@evenhead\@empty
	\def\@oddfoot{}%
	\let\@evenfoot\@oddfoot}
\makeatother

\journal{Mathematics and computers in simulation}

\begin{document}

\begin{frontmatter}



\title{Instability And Network Effects In Innovative Markets}


\author[label1,label2]{P. Sgrignoli}		\ead{paolo.sgrignoli@imtlucca.it}
\author[label3,label4]{E. Agliari}			\ead{elena.agliari@fis.unipr.it}
\author[label3,label4]{R. Burioni}			\ead{raffaella.burioni@fis.unipr.it}
\author[label5]{A. Schianchi}				\ead{augusto.schianchi@unipr.it}

\address[label1]{Università di Verona, Dipartimento di Scienze Economiche, Via dell?Artigliere 19, I-37129 Verona, Italy}
\address[label2]{IMT Institute for Advances Studies Lucca, P.zza San Ponziano 6, I-55100 Lucca, Italy}
\address[label3]{Università di Parma, Dipartimento di Fisica, Viale G.P. Usberti 7/A, I-43100 Parma, Italy}
\address[label4]{INFN, gruppo collegato di Parma, Viale G.P. Usberti 7/A, I-43100 Parma, Italy}
\address[label5]{Università di Parma, Dipartimento di Economia, Via J. Kennedy 6, I-43100 Parma, Italy}

\begin{abstract}
We consider a network of interacting agents and we model the process of choice on the adoption of a given innovative product by means of statistical-mechanics tools. The modelization allows us to focus on the effects of direct interactions among agents in establishing the success or failure of the product itself. Mimicking real systems, the whole population is divided into two sub-communities called, respectively, Innovators and Followers, where the former are assumed to display more influence power. We study in detail and via numerical simulations on a random graph two different scenarios: \emph{no-feedback} interaction, where innovators are cohesive and not sensitively affected by the remaining population, and \emph{feedback} interaction, where the influence of followers on innovators is non negligible. The outcomes are markedly different: in the former case, which corresponds to the creation of a niche in the market, Innovators are able to drive and polarize the whole market. In the latter case the behavior of the market cannot be definitely predicted and become unstable. In both cases we highlight the emergence of collective phenomena and we show how the final outcome, in terms of the number of buyers, is affected by the concentration of innovators and by the interaction strengths among agents.
\end{abstract}


\end{frontmatter}

\section{Introduction}\label{sec:intro}

During the recent years, along with the growing diffusion of new methods of communication, social networking and people direct interactions have been increasingly analyzed by economic research, with new interesting results.

In particular, as stressed in \cite{bib:jack}, a new aspect which emerged is that the classical hypothesis of atomic agents has to be updated in order to allow for  interactions among the individuals themselves. As emphasized in the classic study by Katz and Lazarsfeld \cite{bib:katz}, mass communication needs to rely on individuals to work because they do constitute the basis of the information network and the one-another-influence of decision-makers is a key process in information diffusion \cite[see][]{rogers2003diffusion,delre2010will,bib:Durlauf,bib:banerjee,bib:bik,bib:wied,agliariPRE}. Additionally, what came out from their research (that has been drawn on and improved by Galeotti and many others, see \cite{bib:feick,bib:galeottiFew,bib:galeottiInf,bib:net}) is that the influence process on a community is largely determined by population heterogeneity \cite{young2009innovation,bohlmann2010effects}: information is mediated \emph{by} mass media \emph{to} opinion leaders (or Market Mavens) and \emph{by} them \emph{to} other different social classes, through inter- and intra-groups communication networks. In other words, we deal with a society stratification where each stratum has a different role, from collecting information to creating the network of relations the information will be spread through. Without losing much generality and for the sake of clarity, in most works on this topic, stratification has been simplified in just two classes, often called \emph{Innovators}, i.e. the opinion leaders, and \emph{Followers}, i.e. anyone else \cite{bib:bass,cerqueti}.

In this paper we want to highlight how such inhomogeneity and direct communication can affect the sales performance of a given innovative product. Hence, we consider a market composed by Innovators and Followers who can influence each other through direct (not market mediated) interactions \cite{bib:unidim}; as a result of such interaction they decide whether to buy or not to buy the product. In order to account for the different nature of agents we assume that the interaction strength, that is the influential power, depends on the agents involved. In particular, Innovators, being trend setters and displaying large cohesiveness, will be associated to a higher interaction strength. We also notice that, the degree of Innovators' leading role may depend on the nature of the innovative product considered. Indeed, we distinguish between two possible scenarios.

In the former, referred to as \emph{no-feedback} scenario, the innovativeness of the product is sharp (e.g. determined by remarkable technological improvements) and easily identifiable by consumers; these points make the product non comparable with any other one available. Also, this kind of innovative product is typically characterized by a scarce reachability, due e.g. to high prices, limited number or poor spreading, in such a way that only a part of the population (i.e. Innovators) can afford to buy it and a market niche is established. Under these circumstances Innovators are very cohesive and not prone to abandon the novelty, being negligibly affected by the orientation of the remaining market. On the other hand, Followers, not having direct access to the product are significantly influenced by Innovators and, eventually, attracted by possible discounts, start to acquire it. As an example we can think of Apple Inc.: as they always try to create very original products, they establish brand new market segments. In this scenario we show that Innovators may act as a traditional advertisement, cost-free for the producer and whose effectiveness is directly related to the influence exerted; if the influence is strong enough, Innovators can lead most of market to follow their opinion.

In the other scenario, referred to as with \emph{feedback}, the innovation introduced can be easily and quickly reproduced, in such a way that other brands can produce analogous items. Under these conditions everybody can try the novelty and mutually influence each other. Hence, in this case, the influence on Innovators due to Followers can be non negligible and the former, although being initially buyers, may change their mind. For example, this is the case of functional food products: some consumers do not believe that the innovation is really worth to be paid for and, moreover, the innovation can be so simple (e.g. adding vitamins) that after a short period all the main producers have adopted it. Our results suggest that such situations lead to an unpredictable market behavior, where reproducibility of product failure or success is unlikely, except for borderline cases.

\begin{figure}
\centering
\includegraphics[width=0.9\textwidth]{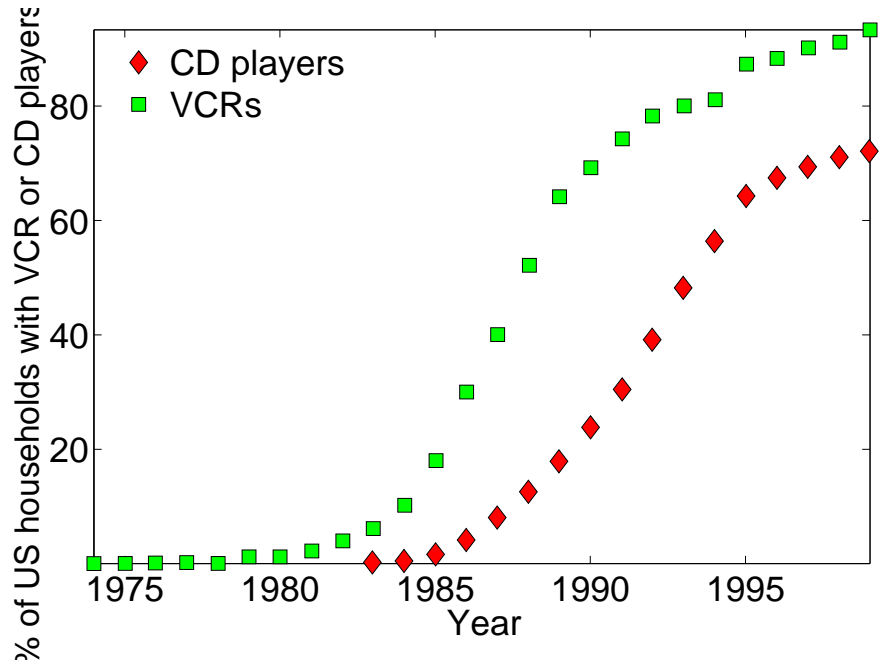}
\caption{Diffusion of CD-ROM and VCR players over about 30 years. Both products experience a rapid growth in diffusion at a given time where the diffusion rate is accordingly much steeper. Collective phenomena (e.g. imitative interactions among users) typically underlie this kind of behavior.}
\label{fig:cd_vcr}
\end{figure}

In both scenarios, we observe the emergence of \emph{collective phenomena}, leading to a global orientation of the market, which is typical of social networks. As a real world example look at figure \ref{fig:cd_vcr}, where the historical data about the diffusion of CD-ROM and VCR players are shown: for both products there is a certain point in time where the market abruptly polarizes and the share of households undergo a steep increase. This kind of behavior is indeed what is called a collective phenomenon and it is a well studied property of ferromagnetic systems, by which our model has been inspired.
We will also show how the final outcome, in terms of the number of buyers, is affected by the concentration of innovators and by the interaction strengths among agents and we will highlight the existence of a critical region in our parameter space where the market is particularly sensitive to small changes and such information could be very useful for market forecasts.

The paper is organized as follows: in sections \ref{sec:motivation} and \ref{sec:model} we introduce our model, its mathematical description and make some remarks on the idea of modeling social phenomena with statistical mechanics tools; in section \ref{sec:numerics} we present our numerical results for the two scenarios and compare the related outcomes. The last section (sec. \ref{sec:conclusions}) is devoted to conclusions and final remarks.

\section{Model motivation} \label{sec:motivation}

In this work we are considering a population where each individual has to decide between two alternatives and his/her behavior is explicitly affected by the previous decisions of his/her acquaintances.

This setting is motivated by many known examples in social sciences where decision makers influence one another because of limited information on the problem or because of an inability to process the information that is available \cite{schelling1973hockey}. When deciding about what movie to watch, which restaurant to go to or what product to buy, one often has scarce information to evaluate the alternatives and therefore relies on others suggestions, or simply pick up the most common choice among the people he/she interacts with. Even when a large amount of information is available, e.g. to evaluate new technology or a risky financial asset, one might not be able to process such information and, again, take decisions based on his/her neighborhood choices. In other contexts, as the so-called social dilemma or in collective action problems, the payoff of our decision could depend explicitly on what others have decided, being higher as many people take the same choice; or in a technology diffusion process the utility of a single additional unit may depend on how many of them are already present in the market, i.e. \emph{network effects}.

In all these kinds of problem individuals have strong incentives to look at others' decisions.

Also the dichotomy of the choice, even if it may appear extremely simplistic, can be relevant in a large number of complex problems: for example the creation of a political coalition or a referendum vote are extremely complex problems with many possible outcomes but when the coalition exists or the referendum text has been drafted the decision is essentially a binary one. Similarly when deciding whether to adopt a new technology or to buy or not any product, the factors involved could be many but the final decision can be again regarded as binary.

This entire class of problems is usually referred to as \emph{binary decisions with externalities} \cite{schelling1973hockey}. Across specific problems the details of the binary decision and the origin of the externalities can vary; nevertheless, in many of the applications that have been examined in the economic literature, the decision can be seen as a function solely of the relative number of agents that have been observed taking an alternative over the other \cite{schelling1973hockey}.

Another big issue in many economic or social problems is that the population is often fragmented: it is very common that people do not behave homogeneously with respect to a certain problem and may interact in different ways. A relevant example is product diffusion processes, where it is known that the market is divided in two sub-communities, usually referred to as \emph{Innovators} and \emph{Followers} \cite{bib:bass}. These two groups display different attitudes with respect to the adoption of novelties and interact differently, depending on whom they are interacting with. Innovators, because of their - by definition - more influential behavior, are also usually called the \emph{trend setters} of the society. This is the kind of scenario we will keep in mind while building our model in the next section.

In general, one can think of many different models of society ranging from purely homogeneous, where each agent has the same number of acquaintances, all of the same kind; to inhomogeneous, where, for example, the number of acquaintances is a random variable and individuals display different attitudes. Such social structures can be effectively envisaged by means of graphs whose nodes represent agents and links between them the existence of a relationship, which could be acquaintanceship, friendship, kinship, etc.
Several kinds of graph have been proposed in the past as able to mimic the features displayed by a real population. In particular \emph{random graphs}, introduced by Erd\"{o}s and R\'{e}nyi \cite{bib:rangraph}, in spite of not being considered to be the most realistic models of real-world networks \cite{strogatz2001exploring}, are often used as their first approximation for combining a stochastic character with an easy tractability that allows to calculate exactly many interesting quantities \cite{jeong2000large,callaway2000network,newman2001random}.

In the following we will adopt this type of graph and, differently from previous works \cite[e.g.][]{watts2002simple}, the dual community structure introduced above.


\section{Model description} \label{sec:model}

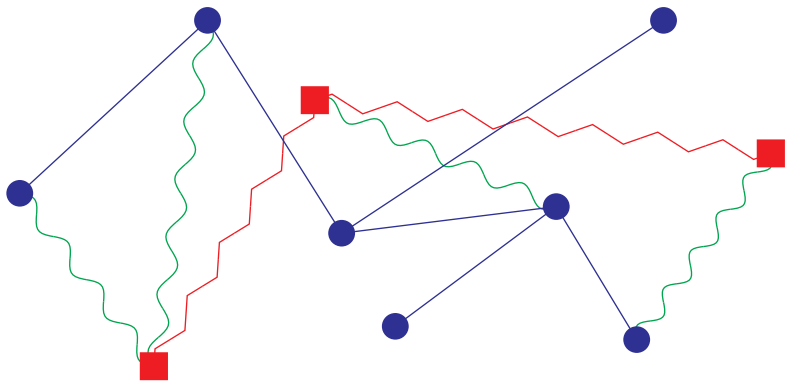
\begin{figure}[htb]
\begin{picture}(300,150)(-50,0)
\SetColor{Green}
\Photon(0,75)(50,10){3}{4}
\Photon(70,140)(50,10){3}{6}
\Photon(110,110)(200,70){3}{5}
\Photon(280,90)(230,20){3}{5}

\SetColor{Red} 
\CBoxc(50,10)(10,10){Red}{Red}
\CBoxc(110,110)(10,10){Red}{Red}
\CBoxc(280,90)(10,10){Red}{Red}
\ZigZag(110,110)(50,10){3}{5}
\ZigZag(280,90)(110,110){3}{7}

\SetColor{Blue} 
\Vertex(0,75){5}
\Vertex(70,140){5}
\Vertex(120,60){5}
\Vertex(140,25){5}
\Vertex(200,70){5}
\Vertex(230,20){5}
\Vertex(240,140){5}
\Line(0,75)(70,140)
\Line(70,140)(120,60)
\Line(120,60)(240,140)
\Line(120,60)(200,70)
\Line(140,25)(200,70)
\Line(200,70)(230,20)
\end{picture}
\caption{Example of a very small graph. Blue dots represent Followers and red squares exemplify Innovators. The first are linked with one another by blue straight lines (tuned by $J_{FF}$) while the second community is connected by red zigzag lines ($J_{II}$); the green wavy lines link Innovators and Followers together ($J_{IF}$ and $J_{FI}$). Each agent $i$ corresponds to a status $\sigma_i$ and feels an external influence $h_i$. Bonds are uniformly drawn with probability $p$. See section \ref{sec:model} for details.}
\label{fig:strutt2pop}
\end{figure}

We consider a society corresponding to a set $I$ of $N$ individuals divided in two different subsets which we identify as \emph{Innovators} ($I_I$) and \emph{Followers} ($I_F$), such that $I = I_I \cup I_F$.

In particular, we define $N$ nodes labeled as $i=1,...,N$ such that $i \in I_I$ for $i = 1,...,N_I$ (Innovators) and $i \in I_F$ for $i = N_I+1,...,N$ (Followers), with $I_I \cup I_F = I$ and $N_I+N_F=N$ and therefore $N=|I|$, $N_I=|I_I|$ and $N_F=|I_F|$ (where $|\cdot|$ represents the number of elements in the set). For any couple $i,j$ we establish a connection with independent probability $p$. We call $z_i$ the degree of node $i$, i.e. the number of agents connected with $i$. Given the way we build the graph, the distribution of $z$ is known to be a binomial \cite{bib:rangraph}. Figure \ref{fig:strutt2pop} represents an example of our model: agents, represented by dots (Followers) and squares (Innovators), interact whenever there exists a link between them. The weight of each link depends on the kind of agents the link is connecting: in the figure, different weights are represented by different line styles.

Each agent of the system has to make a discrete binary choice concerning the adoption of an innovative product introduced in the market \cite[see][]{bib:libro2,bib:libro}, that is each agent can either buy or not buy the product. The state of each agent is therefore encoded by a binary variable denoted as $\sigma$, e.g. $\sigma = +1$ ($\sigma = -1$) means that the agent is (not) a buyer.

The state of any agent is the result of the interactions he/she experiences within the society through, for example, word-of-mouth, e-mail exchange or pure imitation. Moreover the interaction strength between two agents depends on the parts involved: we introduce four parameters $J_{II}$, $J_{FF}$, $J_{IF/FI}$ which represent the coupling strength between two Innovators, between two Followers and between an Innovator and a Follower respectively (see Sec. \ref{sec:Ising}).
In our framework one will expect to have high values of $J_{II}$, small values of $J_{FF}$ and $J_{IF} > J_{FI}$, respectively due to the facts that Innovators are very cohesive, Followers have rather independent behavior and that the influence of Innovators on Followers ($J_{IF}$) is, by definition, stronger than the vice-versa.

In addition to the mutual interaction between agents, also called \emph{herd effect}, it is possible to add a \emph{news effect} ($h$) on the population: this will be able to bias the community towards a given choice and, in the products diffusion scenario we are keeping as reference, can be interpreted as an external advertisement; since these are often targeted, its effectiveness will in general depend on the individuals it is applied to.

In the following we are going to consider only \emph{imitative} interactions, as they are shown to be predominant in several social contexts \cite{bib:imi}. As a result, agents tend to follow their acquaintances decisions and to avoid disagreement, therefore an individual can only tend to imitate his/her neighbors or to follow the advertisement message.
For instance, when deciding whether to buy or not to buy a given product, an agent (being he/she an Innovator or a Follower) will look at the neighborhood choices and/or advertising messages and tend towards the same buying behavior; of course, as mentioned above, the degree of influence is larger when the neighborhood is made of Innovators and an analogous scheme applies for advertisement. In a sense, our model is a ``glass society", where the information about the choice is  visible to all the people in contact and each agent takes his decision by  weighting his/her neighbors' choices. This model corresponds quite well to the web communities case, where the preferences of the members are in general shared with all the friends and are immediately visible. Conversely in traditional markets the information can be transferred using a direct and explicit communication as, for example, word-of-mouth, phone calls or mail exchanges.

To summarize, in our world agents are labeled as either Innovators or Followers, both have to decide whether to adopt an innovative product and they are influenced by:
\begin{itemize}
	\item \emph{nearest-neighbors interactions}, each individual tend to
	imitate its neighbors' decisions
	\item \emph{advertising}, that acts on the population as an external force and
	can bias people to take a specific choice.
\end{itemize}
In any case the effectiveness of the interaction depends on the nature of the
agent(s) considered. We aim to study the different effects and strengths of these two kinds of influence on people behavior, focusing on the context of products adoption.

\subsection{The Hamiltonian as a cost function} \label{sec:Ising}

Given a set of $N$ agents arranged according to a given topology and the assumptions discussed above,
the whole system configuration is given by $\{ \sigma \} \equiv \{\sigma_1, \sigma_2,...,\sigma_N \}$ and its status can be described by means of a \emph{cost function} or \emph{Hamiltonian} $H(\{ \sigma \})$, which reads off as \cite{bib:libro2,bib:libro}
\begin{equation} \label{eq:Hamiltonian}
H_N(\{\sigma\},\mathbf{J})=-\frac{1}{N} \sum_{\substack{i,j=1 \\ i \sim j}}^N J_{ij}\sigma_i\sigma_j-\sum_{i=1}^N h_i\sigma_i,
\end{equation}
where the first sum runs over all couples of connected nodes $(i,j)$.

In general a Hamiltonian function implicitly contains a complete description of the system: solving it would lead us to a full characterization of the model configuration-space and to its evolution equations\footnote{However in the case of eq. \ref{eq:Hamiltonian} as in many other interesting cases, there is no simple analytical solution. This is the reason why for many models, as for ours, the resort to numerical simulations becomes necessary.}. Moreover, if $\sigma_i$ do not explicitly depend on time and $h_i$ are scalars (as in our case), the Hamiltonian also represents the total energy of the system.

By using eq. \ref{eq:Hamiltonian} to describe our model, we are already defining the kind of interactions that can take place in our system: the former term refers to direct influence between two agents, while the latter to an external force, $h$, that can be exerted on any single node.
Furthermore these two different interactions ($J_{ij}$ and $h_i$) depend on the nature (either Innovator or Follower) of the agents $i$ and $j$ considered and, recalling that in our model the population is divided in two groups, we can label agents in such a way that the coupling $\mathbf{J}$ is a four block matrix and the external field $\mathbf{h}$ is a vector, given by

\begin{displaymath}
\mathbf{J}=
\begin{array}{ll}
\quad
\overbrace{\qquad}^{N_I}
\overbrace{\qquad \quad \quad}^{N_F}
\\
\left(\begin{array}{c|ccc}
\mathbf{J_{II}} &  & \mathbf{J_{IF}}
\\
\hline
&  &  &
\\
\mathbf{J_{FI}} &  & \mathbf{J_{FF}}
\\
&   &   &
\\
\end{array}\right)
\end{array}
\!\!\!\!\!
\begin{array}{ll}
\\
\left\} \begin{array}{ll}
\\
\end{array}  \right. \!\!\!\!\! N_I
\\
\left\} \begin{array}{ll}
\\
\\
\\
\end{array} \right. \!\!\!\!\! N_F
\end{array}
\;\;\;\;\;\;\;\;
\mathbf{h}=
\left(\begin{array}{ccc|c}
h_{I}
\\
\hline
\\
h_{F}
\\
\\
\end{array}\right)
\!\!\!
\begin{array}{ll}
\left\} \begin{array}{ll}
\\
\end{array}  \right. \!\!\!\!\! N_I
\\
\left\} \begin{array}{ll}
\\
\\
\\
\end{array}  \right. \!\!\!\!\! N_F
\!\!\!\!\!\!
\end{array}
\end{displaymath}

The assumption that only imitative behavior is taking place means that every entry of the matrix $\mathbf{J}$ has to be non-negative, in this way a configuration where two nearest-neighbors share the same status is more favorable\footnote{In the sense that it has a lower energy, as it is clear by substituting in eq. \ref{eq:Hamiltonian}.}. As mentioned above, the Hamiltonian represents a \emph{cost function}; this means that, following statistical mechanics prescription, a change in the state of any agent ($\sigma_i = -1 \rightarrow +1$ or vice versa) is realized through a stochastic process which is more likely the smaller its cost.
We therefore adopt a local dynamic, where the likelihood of a change in the state of a generic agent depends on the pertaining local cost: a flip $\sigma_i \rightarrow \sigma_i' = -\sigma_i$ for the $i$-th agent corresponds to an energy difference of
\begin{equation}
\Delta H_i(\{\sigma\},\sigma'_i,\mathbf{J}) = \frac{\sigma_i}{N}  \sum_{\substack{j=1 \\ j \sim i}}^N J_{ij} \sigma_j + 2 \sigma_i h_i.
\end{equation}

The value of $\Delta H_i$ measures the cost afforded by the system to perform the flip: when all neighbors $\sigma_j$ are aligned and in agreement with $\sigma_i$ the cost is large, especially when the corresponding couplings $J_{ij}$ are large (i.e. when neighbors are Innovators); vice versa, when $\sigma_i$ disagrees with the overall neighborhood, $\Delta H_i$ is negative and the flip is expected to be easier. Thus, in general, the lower is $\Delta H_i$, the more likely is the flip.

Formally we will make the system evolve by means of Monte Carlo (MC) simulations \cite{bib:MC} where, given the magnetic configuration $\{\sigma\}$, the spin-flip $\sigma_i \rightarrow \sigma_i' = - \sigma_i$ on the $i$-th site, extracted randomly, is accepted with probability
\begin{equation} \label{eq:glauberProb}
p(\{\sigma\},\sigma'_i,\mathbf{J})=\frac{1}{1+e^{\Delta H_i(\{\sigma\},\sigma'_i, \mathbf{J})}}.
\end{equation}
Therefore the algorithm to make the system evolve is composed of two different parts: the choice of the node to update, for which we used a uniform distribution over all the agents, and the probability of the spin-flip, for which we used eq. \ref{eq:glauberProb}. The reason why we made this choice is that this is a well-studied dynamics and it is known, when the interaction pattern is homogeneous, i.e. $J_{ij}=k \; \forall i,j \in I$, for driving the system to a well-defined stationary state \cite{bib:MC,bib:2pop}.
Hence we tested its behavior when the coupling $J_{ij}$ is a block matrix with non-negative values and we verified that also in our case this dynamics leads the system to stationary states, with well-defined proprieties that depend on $\mathbf{J}$ and on the concentration of Innovators, defined as $c \equiv N_I/N$, and with fluctuations that scale like $N^{1/2}$

It is worth noting how the flip does not depend on general system proprieties, but generates from the local status of a few nodes, i.e. the selected node itself and its neighbors. Moreover eq. \ref{eq:glauberProb} makes now clear the discussion about the ease of flipping: the lower the cost, i.e. $\Delta H_i$, the higher will be the probability of flipping and the external parameters $J_{ij}$ regulates the chances of energetically unfavorable status changes to happen.

The external field $h$ enters the equation as a drift term that does not depend on the agent's neighbors but just on its own status and that can, as introduced before, bias the flipping probability independently of other external parameters and the overall system configuration.

We also define the observable
\begin{equation}\label{eq:mag}
M(\{\sigma_i\})=\frac{1}{N}\sum_{i \in I}\sigma_i\ ,
\end{equation}
which provides information about the percentage of the market which has oriented towards a given choice. For instance, if a certain parameters configuration drives the system to have the $85\%$ of agents to buy the product, we will have $M=\frac{1}{N}\left( 0.85N - 0.15N \right) = 0.70$. Evidently $M$ displays upper and lower bounds, namely $+1$ and $-1$ and the same applies separately for $M_I$ or $M_F$, representing the state of the two sub-markets.

The resulting model presents some non-trivial properties related to the collective behavior of its constituents. For example, as shown in \cite{bib:2pop}, the model naturally recovers some realistic phenomena such as a logistic growth for the number of buyers and tipping points. More precisely, given a symmetric $\mathbf{J}$, i.e. $J_{IF}=J_{FI}$, being $J_{II}$ and $J_{FF}$ both small, for a fixed percentage of Innovators, if we increase the inter-community communication we observe that there exists a critical value $J_{IF}^c$, at which the number of agents sharing the same status (e.g. buyer/non-buyer) abruptly grows. This behavior constitutes a genuine collective phenomenon due to the intrinsic communication among agents.
In the present work we aim to analyze how the system behaves under more complex interaction patterns that represent a better modelization of realistic social systems.

Finally we summarize the main points introduced in this section:
\begin{itemize}
\item we model the society by means of an Erd\"{o}s-R\'enyi random graph, in such a way that each agent $i$ has a random number of acquaintances with whom exchanging	information
\item agents making up the society $i=1,....,N$ are divided into two groups, Innovators and Followers; the interaction strength between a couple of agents depends on the kind of agents involved
\item each agent is endowed with a dichotomous variable $\sigma$, which	specifies the status (buy/not to buy) of the pertaining agent
\item we introduce a cost function describing the \emph{cost} and, ultimately, the likelihood, of a given configuration $\{ \sigma_1, \sigma_2,..., \sigma_N \}$
\item we defined a {\em dynamic}, namely a set of rules according to which the state of any agents can be modified.
\end{itemize}

\section{Results and discussion} \label{sec:numerics}

Our analysis is based on numerical simulations performed with Monte Carlo algorithms, where the dynamic introduced in the previous section allows to attain a stationary state. Then, the average value of observables such as $M_I$ and $M_F$ is measured and its dependence on system parameters as well as on the initial configuration is investigated. This allows us to obtain both phase diagram of the system, namely to distinguish the regions in the parameters space where the model displays a \emph{paramagnetic} (i.e. a regime where agents act independently) or \emph{ferromagnetic} (i.e. a regime where agents act collectively) behavior. The choice of focusing on the stationary state follows from the observation that in many situations, ranging from polls to marketing analysis, relevant global parameters describing the behavior of large but finite subsamples of the populations are not rapidly changing on the time scale considered: if the results of the experimental measurements are stable, then it can be meaningful to analyze the social system by looking at its equilibrium or stationary behavior.

In our numerical simulations we will reproduce a double-community market with Innovators and Followers, where, as traditionally assumed and already discussed in section \ref{sec:model}, the former are very cohesive, while the latter are more independent. We want to reproduce likely market conditions just after an innovative product launch, hence we will assume that the initial status is $M_I=+0.95$ and $M_F=-0.90$: Innovators have already and uniformly decided to adopt it, while Followers are still reluctant about it\footnote{These initial values of $M_I$ and $M_F$ correspond to the assumption that $97.5\%$ of the Innovators and only $5\%$ of the Followers have already decided to buy the product under consideration.}. We want to figure out the market conditions that will make the system eventually end up in a buyers or non-buyers predominance, as to say a launch success or failure.

Here, we concentrate on two scenarios: a \emph{feedback} and a \emph{no-feedback} one. As we are going to show in the next subsections, the related emerging behaviors are markedly different.

Before proceeding we underline that the estimate of a given observable such as $\langle M \rangle$ is taken to be the average over a number of $10^3$ decorrelated states of the system, once the equilibrium regime has been reached and that the thermalization time and the decorrelation time are taken to be on the order of $10^2$ MC steps. Moreover, the thermal averages obtained are further averaged over different ($\sim 100$) realizations of the underlying structure, having fixed the number of agents, the average coordination number ($p$) and the populations relative width ($c$), in order to account for the stochasticity of the graph; however, in general, statistical errors due to the 'topological average' are significantly smaller than those arising from the thermal average.

\subsection{No-feedback scenario} \label{sec:JnoFB}

We want to construct a market with no external advertisement where $J_{IF}$, i.e. the influence felt by Innovators due to Followers, is weak, while $J_{FI}$, i.e. the influence felt by Followers due to Innovators, is gradually tuned in order to figure out how the system configuration depends on the gap between the two.
This specific case is likely to occur when the product under consideration creates a clear niche in the market: its innovation is recognized as really powerful so that it establishes a new market segment. In this condition it is natural to think that Innovators adopt the product and hardly change their mind about it.
The interaction parameters are defined as follows:
\begin{eqnarray}
J_{IF}&=& J \nonumber \\
J_{FI}&=& J+\gamma \;,
\end{eqnarray}
where $J$ is a finite, fixed parameter and $\gamma$ is tunable and represents the interaction growth. Notice that the situation is asymmetric.

\begin{figure}
\centering
\includegraphics[width=0.75\textwidth]{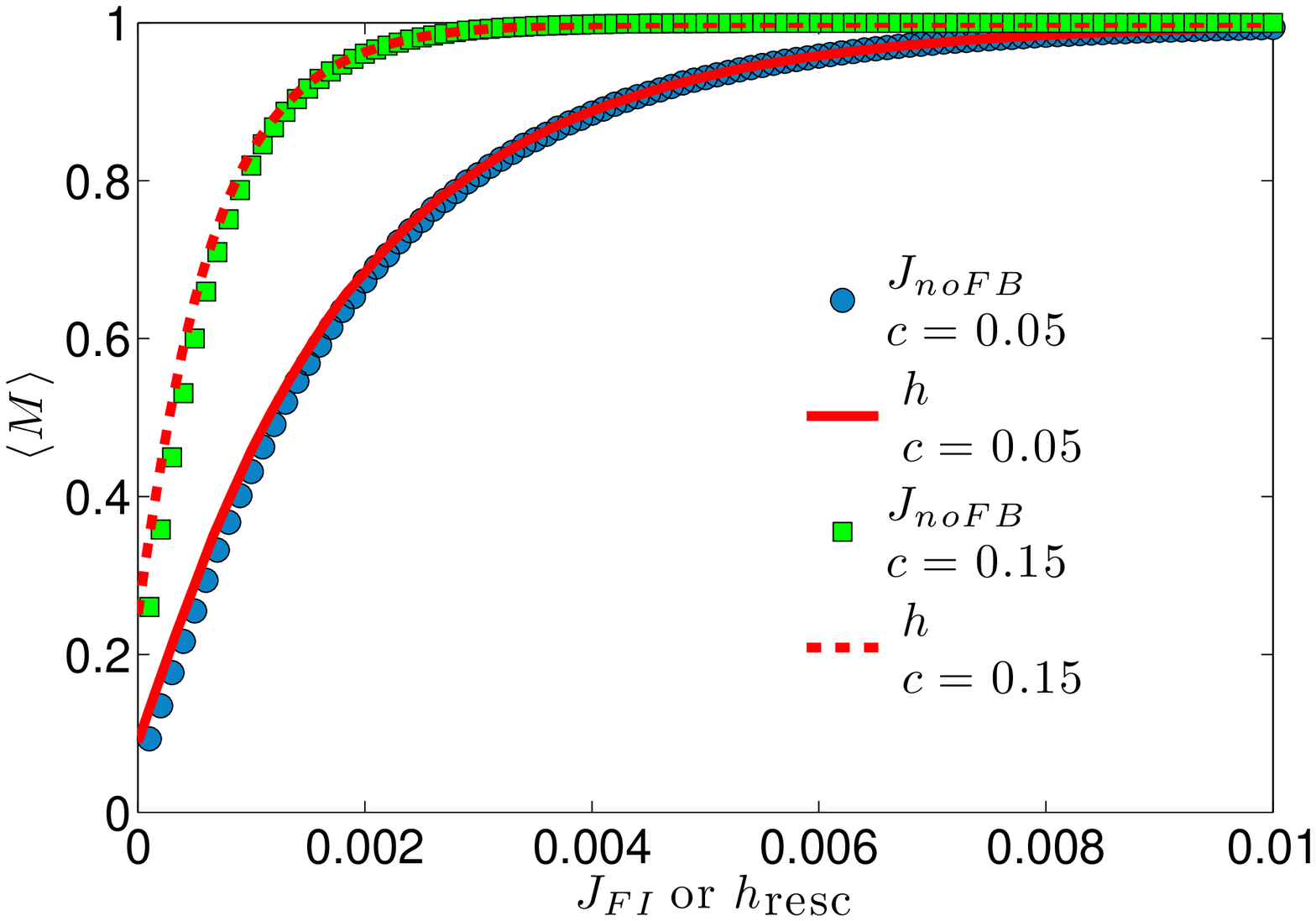}
\includegraphics[width=0.75\textwidth]{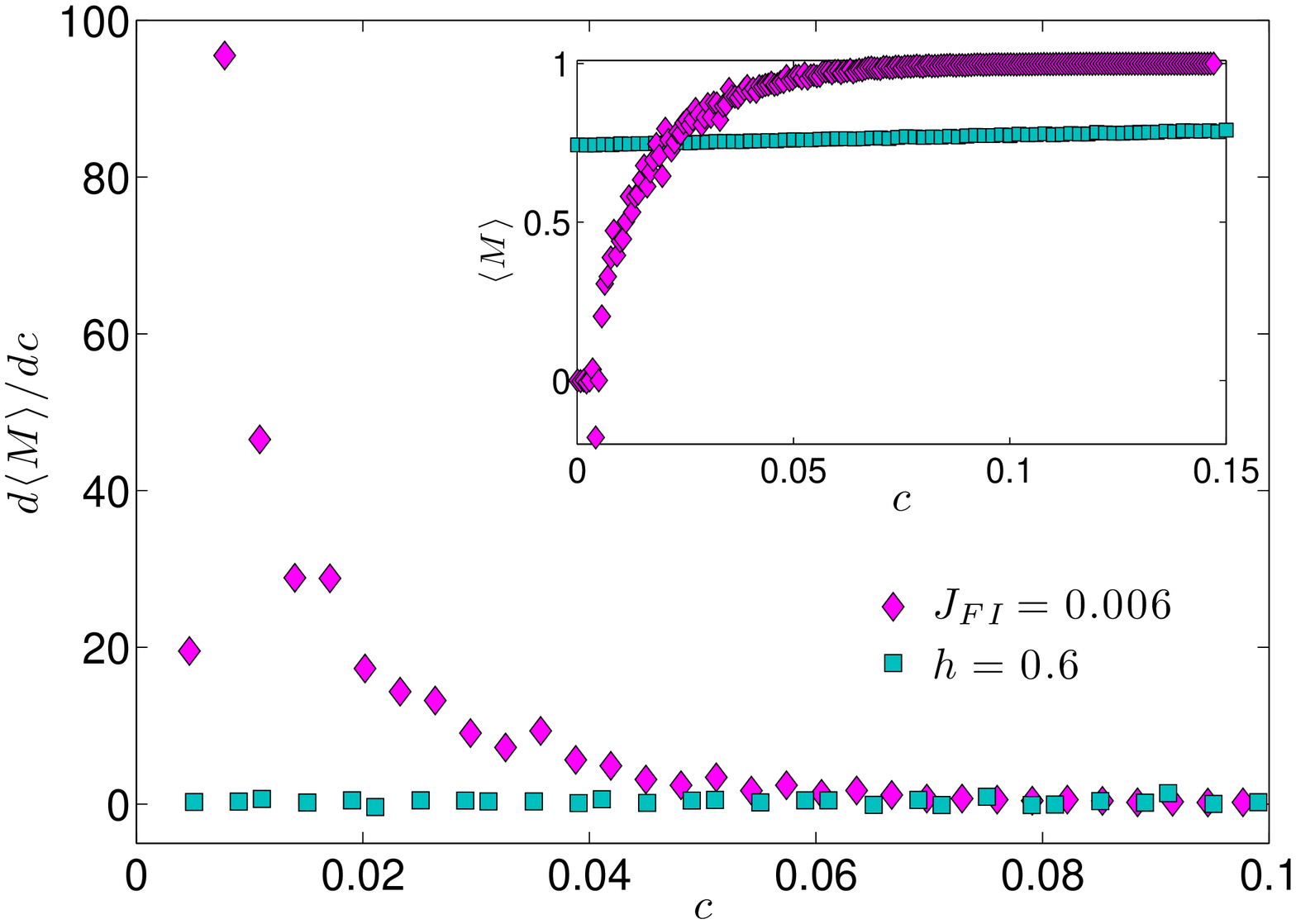}
\caption {Upper panel: the total market choice, when a stationary state is reached, for two values of $c$. Symbols represent data points obtained by varying $\gamma$ and assuming $h=0$, while curves represent data obtained by varying $h$, properly rescaled with $Npc$ for the comparison. Lower panel: market indicator ($\langle M \rangle$) derived with respect to $c$ for both cases. In the inset the data for $\langle M \rangle$ are plotted. Simulations are performed with $N=6000$ agents, $p=0.80$, $J_{IF}=J_{FF}=10^{-4}$ and $J_{II}=0.05$.}
\label{fig:jnofb}
\end{figure}

In Figure \ref{fig:jnofb} (top panel) we can see how the system responds as $\gamma$ grows, compared with data from a case with widespread advertisement and no mutual interaction among agents, that is $\gamma=0$ and $h \neq 0$. The two cases are almost overlapping, meaning that Innovators act like an external field on the remaining system and indeed we observe a similar growth rate; their cohesive and buying-oriented behavior, without a significant feedback, make them act as an effective forcing on the whole system.

It is very important to notice that in Figure \ref{fig:jnofb} (top panel), in order to compare data sets from the two cases, we had to rescale the advertisement magnitude: the idea is that \emph{each} Innovator in our system acts like an external field on his/her neighborhood and, since we set up a world with a high density of connections (i.e. high $p$), they can actually have a big influence on the whole system. Hence to compare data we divided $h$ by the average number of Innovator neighbors every agent has, i.e. $h \rightarrow \tilde{h}=\frac{h}{Npc}$. This feature of our model tells us that acting on Innovators visibility, in a market that is likely to have low to none feedback, we can significantly boost our sales: it can be sufficient giving any Innovator a weak influence/visibility power, e.g. viral ads, to obtain market shares at least as strong as with canonical advertisement but with evident lower costs than TV or magazines ads.

Furthermore, we can study the dependence displayed by the market indicator $\langle M \rangle$ on $c$, in order to understand how and when acting on the Innovators percentage can be useful for sales. In figure \ref{fig:jnofb} (bottom panel) we can see the quantity $\partial \langle M \rangle/ \partial c$, that indicates how market share changes depending on a small variation of the percentage of trend setters in the community. As it is clear from the figure, the effect due to agent communication and to widespread advertisement are significantly different: in the former case $\partial \langle M \rangle/ \partial c$ shows a peak for low values of $c$ ($\sim 1\%$), while the latter is close to zero value on the entire range analyzed. This means that, especially for small values of $c$, a tiny increase in the Innovators number causes a dramatic variation in the share of buyers in the market (over $90\%$), while for the news case the two quantities are unrelated.
These results can be used to decide for the best strategy of a company, assuming we are able to understand the market conditions: if we are relying on viral diffusion of our marketing information, it should be convenient to make investments aiming to increase the Innovators number up to some percentage point (at $3 - 4\%$ we still have gains of $\sim 10\%$) while efforts are almost useless if the reference community already has more than $5\%$ of trend setters.

This whole analysis leads us to conclude that a market like the one we just depicted behaves in a good predictable manner: a sub-community of Innovators that receive no feedback about their choices and cohesively adopt a certain product can eventually lead to the complete polarization of the population. As we will see in the next section, when feedback comes into play this deterministic and ordered behavior ceases to exist.

\subsection{Feedback scenario and market instability} \label{sec:JFB}

\begin{figure}
\centering
\includegraphics[width=0.75\textwidth]{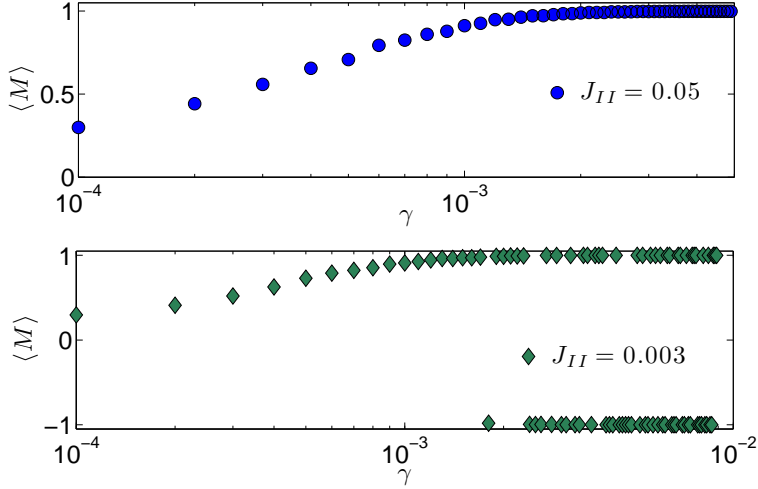}
\caption{The total market choice $\langle M \rangle$ as a function of $J$, for fixed $c=0.15$, while $J_{II}$ is taken equal to $0.05$ (upper panel) and equal to $0.003$ (lower panel). Simulations are performed with $N=6000$ agents, $p=0.80$, $J_{FF}=10^{-4}$ and $\varepsilon=0.5$.}
\label{fig:jfb}
\end{figure}

We want to build a market where, like before, there is no external advertisement and information about new products is driven only by word-of-mouth among agents. However, differently from the previous one, now we allow for feedback interactions: as we increase Innovators vs Followers communication strength ($J_{FI}$), the inverse interaction ($J_{IF}$) gets stronger too. In other words, the influence power always remains asymmetric (i.e. $J_{FI} > J_{IF}$) but when $J_{FI}$ is increased, $J_{IF}$ is also strengthened, so that their ratio remains constant (see eq. \ref{eq:fb}). In this case, any agent has the chance to outline an opinion concerning the product so that the Innovators role is weaker and they can be non-negligibly influenced as well.

More precisely, we consider the following interaction: introduce a new parameter $\varepsilon$ and let
\begin{align} \label{eq:fb}
J_{IF} &= (1-\varepsilon)\ \gamma \;, \nonumber \\
J_{FI} &= (1+\varepsilon)\ \gamma \;, \\
\frac{J_{IF}}{J_{FI}} &= \frac{1-\varepsilon}{1+\varepsilon} \;. \nonumber
\end{align}
Here $\varepsilon$ represents a percentage difference of the interaction in the two directions. In the following simulations we will make the value of $\gamma$ run for different configurations of $c$, keeping, just as an example, $\varepsilon=0.50$ fixed.
It turns out that this setting leads to unpredictable results: for the same parameters configuration, the market may end up in a product success or failure randomly.

In Figure \ref{fig:jfb} we can see the results obtained. We compare the cases for two different values of $J_{II}$ ($0.003$ and $0.05$): we can clearly observe that if Innovators cohesiveness is not strong enough, the system, for higher values of inter-population interactions, became unstable and unpredictable. This is an effect of the feedback that characterizes this scenario. Indeed, as the interaction grows Innovators are more exposed to opinions from the rest of the market that we supposed initially against the new product. On the contrary, a market where Innovators present very strong bonds among one another, gives the same results as the study of Section \ref{sec:JnoFB}.

\begin{figure}
\centering
\includegraphics[width=0.75\textwidth]{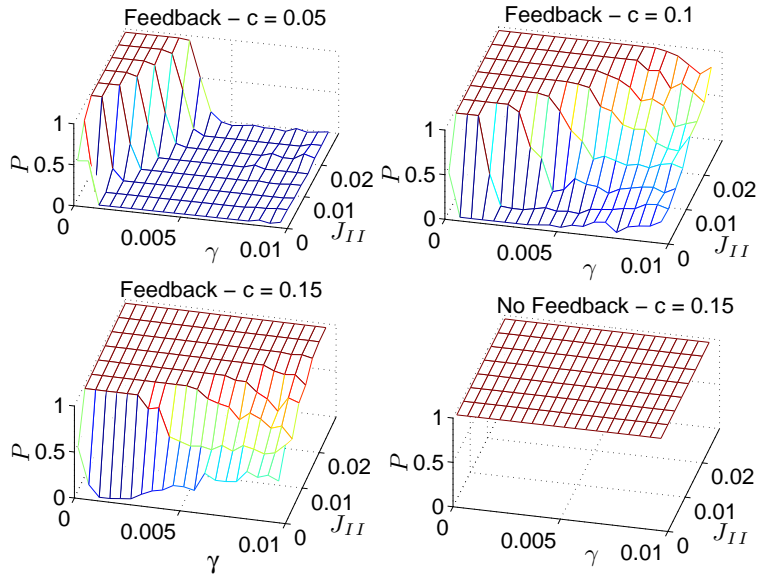}
\caption{Market success probability in the feedback ($\varepsilon=0.50$) scenario as a function of $J_{II}$ and $\gamma$; different panels correspond to different values of $c$. The last panel shows, for comparison, an analogous market with no feedback. All results refer to a system with $N=6000$ agents and $p=0.80$.}
\label{fig:prob}
\end{figure}

The unpredictability is evidently the key-element of our model in this scenario.
In order to deal with it we study the probability of success of the market: without a deterministic behavior what we want to look at is how many times, out of a certain number of tries, a certain market condition leads to a successful product diffusion. In Figure \ref{fig:prob} we represent this probability for the cases $c=0.05,\ 0.10,\ 0.15$ and we observe that both $J_{II}$ and $\gamma$ play important roles: an increase in $\gamma$ increases the feedback as well and makes the success probability undergo a drop from almost $1$ to nearly $0$; in a similar way, for a fixed $\gamma$, there is a value of $J_{II}$ around which the probability sharply changes. Indeed, for low values of $J_{II}$ Innovators are not very connected so that they cannot really lead the market; this is true as far as $J_{II}$ reaches a certain tipping point (that depends also on the value of $\gamma$) where, instead, their action as \emph{trend setters} turns out to be successful in polarizing the market.

The system \emph{instability} we observe originates from a lack of cohesiveness of the leading community: when the two populations are opposed with comparable strength, the outcome of their interaction is no longer a universal and reproducible result but heavily depends on the model's initial specification and local topology. In this case how Innovators and Followers are located and connected one another on the graph become of extreme importance in the establishment of the final opinion of the entire community. This result can be interpreted as a warning for policy makers about the complexity of multiple community scenarios.

\section{Conclusions} \label{sec:conclusions}

In this work we introduced a model for the behavior of a market composed of interacting agents, when an innovative product is introduced. The structure of the society is represented by a random graph, whose nodes represent agents and only neighboring ones are allowed to interact and influence each other. The likelihood that a given agent will adopt the new product depends on the number, state and type of his/her acquaintances. Mimicking real systems, we indeed distinguish between two kinds of agents: Innovators and Followers, characterized by different attitudes towards the new product and different degrees of influence with respect to other individuals.

According to the kind of product under consideration, we distinguish between two scenarios.
In the former case, the product determines a real breakthrough giving rise to a market niche where initially only Innovators access; because of this they will be particularly cohesive and scarcely affected by the remaining of the market. Under such conditions we found that, by tuning Innovators influential power, it is always possible for them to drive market opinions. In other words, they work as a (cost-free) amplified advertisement.

Conversely, when the innovativeness of the product can be easily and quickly reproduced by other firms, it is plausible that Innovators feel a \emph{feedback} due to the remaining population. Innovators no longer constitute a separate market but they are integrated in the community and they can be strongly influenced too. We have shown that this condition gives rise to non-predictable market results: the system may either succeed or fail for the same parameters configuration. This behavior suggests a strong sensitivity of the system about initial conditions.

Hence, according to the kind of product considered, the behavior of the market can be dramatically different, ranging from a well predictable system where the visibility of Innovators has direct impact on the product success, to a non-predictable one, where acting on Innovators influence may or may not yield any significant improvement for sales performance.

Among the possible developments of this work we mention the extension to more inhomogeneous networks as models for the social structure \cite{bib:SW}. Beyond the rather homogeneous random graph adopted here, one could extend the analysis to societies where the node degrees span over a wider range as in scale-free networks: those graphs display a large number of nodes scarcely connected and a few nodes, called hubs, with a significantly large number of neighbors. Whether Innovators are placed on hubs or not is expected to have crucial effects on the overall behavior of the system.

\bibliographystyle{model1b-num-names}
\bibliography{myrefs}

\end{document}